\documentclass[prd,aps,showpacs,preprintnumbers,amssymb]{revtex4}
\usepackage{axodraw}
\usepackage{color}
\usepackage{epsf}

\def\e3p{$\eta \rightarrow 3 \pi$}

\begin{document}
\title{%
\hfill{\normalsize\vbox{%
\hbox{}
 }}\\
{A hint of a  strong supersymmetric standard model }}

\author{Renata Jora
$^{\it \bf a}$~\footnote[2]{Email:
 rjora@theory.nipne.ro}}

\author{Joseph Schechter
 $^{\it \bf b}$~\footnote[4]{Email:
 schechte@phy.syr.edu}}

\affiliation{$^{\bf \it b}$ National Institute of Physics and Nuclear Engineering PO Box MG-6, Bucharest-Magurele, Romania}

\affiliation{$^ {\bf \it d}$ Department of Physics,
 Syracuse University, Syracuse, NY 13244-1130, USA}

\date{\today}

\begin{abstract}
We discuss the supersymmetric standard model from the perspective that the up and down Higgs supermultiplets are composite states. We show
that a Higgs multiplet in which the scalar states are bound states of two squarks and the corresponding Higgsinos  are bound states of a quark and an squark
has the correct supersymmetry transformations and may lead to an alternative model which displays dynamical symmetry breaking. We describe this model through an effective Higgs potential
which by itself may lead to the correct mass of $125.9$ GeV for the lightest Higgs boson and to other neutral scalar masses respecting the experimental constraints.
\end{abstract}
\maketitle

\section{Introduction}
The remarkable results of the LHC experiments \cite{Atlas}, \cite{CMS} culminating with the discovery of a Higgs  boson with a mass $m_h=125.9$ GeV
has limited the parameter space of many beyond the standard model theories \cite{Lykken} without eliminating them completely. Among the most popular candidates of this type are the supersymmetric models with their minimal version the MSSM. The fact that the supersymmetry breaking scale has been pushed further up by the experimental constraints might question
if the naturalness \cite{Arkani} should be one of the main driving reason for low scale supersymmetry. In this context it would be interesting to explore other avenues supersymmetry related like that of a supersymmetric model with strong dynamics. In what follows we shall consider this point of view and sketch a possible picture in this direction. Inspired by the low energy QCD and by models with dynamical symmetry breaking  we shall  hypothesize that the squarks of one generation (assume the top and bottom quarks) form bound states and condense thus leading to both electroweak and supersymmetry breaking.  The corresponding operators exist already in the MSSM Lagrangian but instead on dwelling on them we will consider a simple effective Lagrangian which parameterizes our ignorance of the underlying strong dynamics.

  \section{A strong version of the MSSM}

We start with the MSSM model without the soft supersymmetry breaking terms. We then set the gauge kinetic term plus the gauge interaction terms for the up and down Higgs doublets to zero.
We can do that without altering the supersymmetric nature of the model as  these terms  are invariant under the supersymmetry transformation by themselves.
The superpotential is that of the MSSM, specifically:
\begin{eqnarray}
W_{MSSM}=y_u\tilde{u}^{*}_R\tilde{Q}_u H_u-y_d\tilde{d}^{*}\tilde{Q}_d H_d-y_e\tilde{e}^*_R\tilde{L}H_d+\mu H_u H_d.
\label{superp090}
\end{eqnarray}

For simplicity in what follows we will consider only one generation of the standard model fermions and sfermions denoted generically by up and down.

From the  absence of the F terms for the up and down Higgs doublets we deduce:
\begin{eqnarray}
&&\frac{\partial {\cal L}}{\partial F_u}=-W^{u*}=0
\nonumber\\
&&\frac{\partial {\cal L}}{\partial F^{u*}}=-W_u=0
\nonumber\\
&&\frac{\partial {\cal L}}{\partial F_d}=-W^{d*}=0
\nonumber\\
&&\frac{\partial {\cal L}}{\partial F^{d*}}=-W_d=0.
\label{rez5647}
\end{eqnarray}

This leads to:
\begin{eqnarray}
&&H_u=\frac{1}{\mu}[y_d\tilde{d}^{*}\tilde{Q}_d +y_e\tilde{e}^*_R\tilde{L}]
\nonumber\\
&&H_d=-\frac{1}{\mu}y_u\tilde{u}^{*}_R\tilde{Q}_u.
\label{rez56}
\end{eqnarray}

Replaced in the superpotential this leads to a new expression:

\begin{eqnarray}
W_{MSSM}=\frac{1}{\mu}[y_d\tilde{d}^{*}\tilde{Q}_d +y_e\tilde{e}^*_R\tilde{L}]y_u\tilde{u}^{*}_R\tilde{Q}_u.
\label{superpot56}
\end{eqnarray}

Note that the chiral superfields corresponding to the up and down Higgs doublets have been eliminated completely and we obtain an effective superpotential.

If we neglect the terms of order $\frac{1}{\mu}$ this superpotential will lead to the following Lagrangian:
\begin{eqnarray}
{\cal L} =-y_u(\bar{u}u H_u^0-\bar{u}d H_u^+)+y_d(\bar{d}u-\bar{d}d H_d^0)+y_{e}(\bar{e}\nu_e H_d^--\bar{e}e H_d^0)
\label{lagr567}
\end{eqnarray}
where for simplicity we adopted the notation:
\begin{eqnarray}
&&H_u^0=\frac{1}{\mu}(y_d\tilde{d}_R^{*}\tilde{d}_L +y_e\tilde{e}_R^*\tilde{e}_L)
\nonumber\\
&&H_u^+=\frac{1}{\mu}(y_d\tilde{d}_R^{*}\tilde{u}_L +y_e\tilde{e}_R^*\tilde{\nu_e})
\nonumber\\
&&H_d^0=-\frac{1}{\mu}(y_u\tilde{u}_R^*\tilde{u}_L)
\nonumber\\
&&H_d^-=-\frac{1}{\mu}(y_u\tilde{u}_R^*\tilde{d}_L).
\label{noy768}
\end{eqnarray}

This Lagrangian contains the standard couplings of the Higgs bosons with the standard model fermions.

However since it is hard to justify the formation of a two lepton composite state we shall adopt a picture where the composite Higgses correspond to one up and one down quark such that a dynamical mass will be generated for these.  We introduce the up Higgs supermultiplet where for example the neutral states have the structure:
\begin{eqnarray}
&&H_u^0=\frac{1}{\mu}\tilde{d}_R^*\tilde{d}_L
\nonumber\\
&&\tilde{H}_u^0=\frac{1}{\mu}(d_R^{\dagger}\tilde{d}_L+\tilde{d}_R^*d_L).
\label{superm78}
\end{eqnarray}
All other fermions and sfermions then couple as in the regular MSSM with the up and down composite Higgses.
We further check that these composite states still respect the supersymmetry transformation.
We will show the invariance under the regular supersymmetry transformation only for the neutral parts of the up supermultiplet:
\begin{eqnarray}
&&\delta(H_u^0)=\frac{1}{\mu}(y_d\delta(\tilde{d}_R^*\tilde{d}_L)=
\frac{1}{\mu}y_d[(\delta\tilde{d}_R^*)d_L+\tilde{d}_R^*\delta(\tilde{d}_L)]=
\frac{1}{\mu}y_d\epsilon(d_R^{\dagger}\tilde{d}_L+\tilde{d}_R^*d_L)=
\epsilon \tilde{H}_u^0
\nonumber\\
&&\delta\tilde{H}_u^0=\frac{1}{\mu}[-i[((\sigma^{\mu}\epsilon^{\dagger}_{\dot{\alpha}})\Delta \tilde{d}_R^*)\tilde{d}_L+((\sigma^{\mu}\epsilon^{\dagger}_{\dot{\alpha}})\Delta \tilde{d}_L)\tilde{d}_R^*]+\delta(\tilde{d}_R^*)d_L+d_R^{\dagger}\delta(\tilde{d}_L)]
\nonumber\\
&&=-i\frac{1}{\mu}[(\sigma^{\mu}\epsilon^{\dagger}_{\dot{\alpha}})\Delta (\tilde{d}_R^*\tilde{d_L})]-\frac{1}{\mu}\epsilon (d_R^{\dagger}d_L)=-i(\sigma^{\mu}\epsilon^{\dagger}_{\dot{\alpha}})\Delta(H_u^0)-\frac{1}{\mu}\epsilon (d_R^{\dagger}d_L).
\label{finalrez56478}
\end{eqnarray}
The last term in the second equation of (\ref{finalrez56478}) would correspond to an $F_u$ contribution or better to a $W_u^*$ term associated with a possible superpotential.

Note that the regular MSSM Lagrangian with the elementary states already contains the seeds of possible dynamical electroweak and supersymmetry breaking through the contributions from the D terms:

\begin{eqnarray}
\frac{1}{2}\frac{g^{'2}}{4}[Y_{uL}Y_{uR}\tilde{u}_L^*\tilde{u}_L\tilde{u}_R^*\tilde{u}_R+Y_{dL}Y_{dR}\tilde{d}_L^*\tilde{d}_L\tilde{d}_R^*\tilde{d}_R],
\label{foursq45}
\end{eqnarray}
such that two scalar condensates may form. Here $Y_{L,R}$ are the hypercharges corresponding to each particle.

This is not unnatural for low energy QCD where it is hypothesized that the scalar states are an admixture of two quark and four quark states \cite{Jora}. Moreover the four quark states may be regarded as molecule formed out from two quark states.
For example if the two quark state is given by,
\begin{eqnarray}
M_a^b=(q_{bA})^{\dagger}\gamma^4\frac{1+\gamma_5}{2}q_{aA},
\label{tw5534}
\end{eqnarray}
there is a possibility that the four quark state have the form:

\begin{eqnarray}
M^{'b}_a=\epsilon_{acd}\epsilon^{bef}(M^{\dagger})_e^c(M^{\dagger})_f^d.
\label{four6657}
\end{eqnarray}

Then the low energy dynamics \cite{Jora} suggests that the vacuum expectation expectation values of the scalar mesons and the two and four quark condensates are related as in:
\begin{eqnarray}
&&\alpha=-\frac{1}{2\Lambda^2} \langle \bar{q}_{1A}q_{1A}\rangle=-\frac{1}{2\Lambda^2} \langle M_{11}\rangle
\nonumber\\
&&\beta=-\frac{\omega}{\Lambda^5} \langle M'_{11}\rangle.
\label{rez234}
\end{eqnarray}
Here $\alpha$ and $\beta$ are the vacuum  expectation values for the two quark and four quark scalar states whereas the quantities in the brackets are the two quark and
four quark vacuum condensates respectively. Moreover $\omega$ which is adimensional parameter is determined to be (from the instanton dynamics) in the range $1-20$.
Low energy spectroscopy of scalar and pseudoscalar indicate that $\alpha\simeq \beta$.

Note that the existence of possible two squarks bound states has been explored earlier in \cite{Peccei1}, \cite{Peccei2} with the strong dynamics related to a large squark-squark-Higgs coupling. However here we adopt a picture in which the strong mechanism is similar to low energy QCD and may require the introduction of an additional strong gauge interaction. The resulting effective theory has thus in terms of matter particles only the minimal content of the MSSM with the regular two Higgses.


\section{The scalar masses}

We shall describe a strongly interaction supersymmetric model through an effective theory theory which by itself in the absence of quantum corrections should lead to the correct mass of the Higgs boson found at the LHC. As an illustration we assume that the usual MSSM Higgs potential is generated together with two beyond MSSM effective terms introduced first in \cite{Seiberg}.
One of them is supersymmetric and stems from the effective superpotential,
\begin{eqnarray}
W_e=\frac{1}{M}(H_u^2H_d^2),
\label{eff98}
\end{eqnarray}
the other one is supersymmetry breaking term. Together they bring the following contribution to the Higgs potential:

\begin{eqnarray}
V_1+V_2=2x_1(H_u^2H_u^*H_d+H_d^2H_d^*H_u)+x_2(H_u^2H_d^2)+c.c.
\label{rez6758}
\end{eqnarray}

Here $x_1$ and $x_2$ are small coefficients which parameterize the ratio of the supersymmetry scale to that of possible new physics in general but for our case at hand we consider them
to parameterize the unknown strong dynamics. In particluar $x_1=\frac{\mu}{M}$ where $\mu$ is the usual MSSM scale. For simplicity we shall take the parameters $x_1$ and $x_2$ real although there is no need to do so.

The full Higgs potential:
\begin{eqnarray}
V=V_{MSSM}+V_1+V_2,
\label{rez43567}
\end{eqnarray}
will lead to the minimum equations:

\begin{eqnarray}
&&|\mu|^2+m_{H_u}^2-b\frac{v_d}{v_u}+\frac{1}{4}(g^2+g^{\prime 2})(v_u^2-v_d^2)+6x_1v_uv_d+2x_1\frac{v_d^3}{v_u}+2x_2v_d^2=0
\nonumber\\
&&|\mu|^2+m_{H_d}^2-b\frac{v_u}{v_d}+\frac{1}{4}(g^2+g^{\prime 2})(V_d^2-v_u^2)+6x_1x_2v_uv_d+2x_1\frac{v_u^3}{v_d}+2x_2v_u^2=0.
\label{min786}
\end{eqnarray}

We introduce the notation $\tan\beta=\frac{v_u}{v_d}$. Moreover we express the two Higgs doublets in terms of the mass eigenvalues in the usual way.
\begin{eqnarray}
\left(
\begin{array}{c}
H_u^0\\
H_d^0
\end{array}
\right)=
\left(
\begin{array}{c}
v_u\\
v_d
\end{array}
\right)+
\frac{1}{\sqrt{2}}
R_{\alpha}
\left(
\begin{array}{c}
h\\
H
\end{array}
\right)
+\frac{i}{\sqrt{2}}R_{\beta_0}
\left(
\begin{array}{c}
G^0\\
A^0
\end{array}
\right).
\label{matr6476}
\end{eqnarray}

Here,
\begin{eqnarray}
R_{\alpha}=
\left(
\begin{array}{cc}
\cos{\alpha}&\sin{\alpha}\\
-\sin{\alpha}&\cos{\alpha}
\end{array}
\right),
\label{rot6578}
\end{eqnarray}

and,
\begin{eqnarray}
R_{\beta_0}=
\left(
\begin{array}{cc}
\sin{\beta_0}&\cos{\beta_0}\\
-\cos{\beta_0}&\sin{\beta_0}
\end{array}
\right).
\label{rot568789}
\end{eqnarray}

We shall ignore the charged states in what follows.

The mass matrix for the scalar Higgs boson is  calculated to be:
\begin{eqnarray}
&&(M^2_{h,H})_{11}=m_Z^2\frac{v_u^2}{v^2}+6x_1v_uv_d-2x_1\frac{v_d^3}{v_u}+b\frac{v_d}{v_u}
\nonumber\\
&&(M^2_{h,H})_{22}=m_Z^2\frac{v_d^2}{v^2}+6x_1v_uv_d-2x_1\frac{v_u^3}{v_d}+b\frac{v_u}{v_d}
\nonumber\\
&&(M^2_{h,H})_{12}=-m_Z^2\frac{v_uv_d}{v^2}+6x_1v^2+4x_2v_uv_d-b,
\label{firstm768}
\end{eqnarray}
whereas that of the pseudoscalar states is:
\begin{eqnarray}
&&(M^2_{G,A})_{11}=b\frac{v_d}{v_u}-2x_1v_uv_d-2x_1\frac{v_d^3}{v_u}-4x_2v_d^2
\nonumber\\
&&(M^2_{G,A})_{22}=b\frac{v_u}{v_d}-2x_1v_uv_d-2x_1\frac{v_u^3}{v_d}-4x_2v_u^2
\nonumber\\
&&(M^2_{G,A})_{12}=b-2x_1v^2-4x_2v_uv_d.
\label{os678}
\end{eqnarray}

In order to see if the above mass matrices are compatible with the latest LHC data \cite{Atlas}, \cite{CMS} and especially with a light Higgs boson with a mass at $m_h=125.9$ GeV
 we will use the results of the phenomenological study done in \cite{Polosa}. There it is shown that the best fit points for the ratio of the Higgs signal strenghts (the ratios
 $R_{\gamma\gamma}=\mu_{\gamma\gamma}/\mu_{ZZ}$ and $R_{\tau\tau}=\mu_{\tau\tau}/\mu_{WW}$, $\mu_{XX}=\frac{\sigma(pp\rightarrow hh)BR(h \rightarrow XX)}{\sigma(pp\rightarrow hh)_{SM}BR(h \rightarrow XX)_{SM}}$) is for $\tan\beta\approx 1$ and for $\alpha$ very close to the decoupling limit $\alpha=\beta-\pi/2$.
 We shall use these value together with the mass of the Higgs boson found at the LHC as inputs.

\begin{figure}
\begin{center}
\epsfxsize = 10cm
 \epsfbox{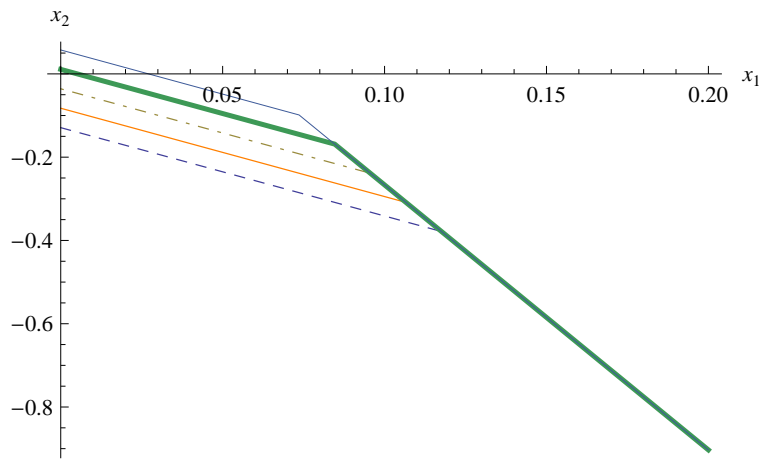}
\end{center}
\caption[]{%
Plot of the parameter $x_2$ (dashed line) as a function of the parameter $x_1$ for five values of the parameter b: 1 TeV (dashed line), 2 TeV (orange line), 3 TeV (dotdashed line), 4 TeV (thick line), 5 TeV (thin line). Here we used the best fit point for $\tan \beta$ found from the ratio of the Higgs signal strengths in \cite{Polosa}.}
\label{s1}
\end{figure}

\begin{figure}
\begin{center}
\epsfxsize = 10cm
 \epsfbox{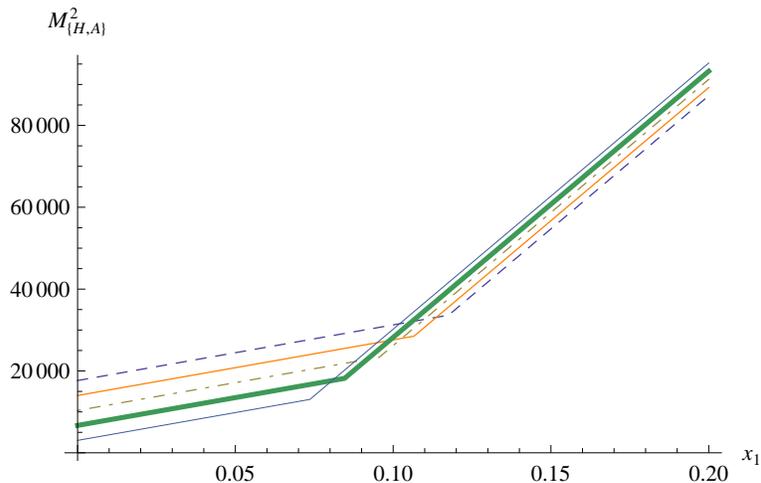}
\end{center}
\caption[]{%
Plot of $M_H^2=M_A^2$ as a function of the parameter $x_1$ for five values of the parameter b: 1 TeV (dashed line), 2 TeV (orange line), 3 TeV (dotdashed line), 4 TeV (thick line), 5 TeV (thin line). Here we used the best fit point for $\tan \beta$ found from the ratio of the Higgs signal strengths in \cite{Polosa}.}
\label{s2}
\end{figure}

 In Fig \ref{s1} we plot the parameter $x_2$ as a function of the parameter $x_1$ for five values of the parameter b in the range 1-5 TeV. In the limit considered the mass of heaviest Higgs boson H and that of the pseudoscalar A are degenerate. In Fig \ref{s2} we plot the square of these masses as a function of the parameter $x_1$ in the same range of masses for b.
 The experimental lower bounds that exist on $M_H$ and $M_A$ agree better with those region of the parameter b is lower in value whereas $x_1$ and $x_2$ are higher in absolute values although still much lesser than 1.

\section{Discussion}

We consider the possibility that the supersymmetric standard model has an underlying strong dynamics where the Higgses are bound states of two squarks whereas the corresponding Higgsinos
are composite states of an squark and a quark. Correct supersymmetry transformations can be preserved for these states. The formation of two scalar vacuum condensates due to the color group or possibly an additional gauge sector may trigger both electroweak and supersymmetry breaking. Since it is hard to describe such an intricate model we parameterize the underlying strong dynamics through the introduction of two effective terms in the scalar potential, one supersymmetric the other one with supersymmetry breaking.  This new potential can lead by itself to the correct mass of the light Higgs boson found at the LHC.

We further show that this picture is consistent with the latest LHC data and the corresponding experimental constraints lead to degenerate masses for the H and A bosons which may take a large range of values, depending on the choice of parameters.

\section*{Acknowledgments} \vskip -.5cm
The work of R. J. was supported by a grant of the Ministry of National Education, CNCS-UEFISCDI, project number PN-II-ID-PCE-2012-4-0078. We  would like to thank Salah Nasri for useful comments about the manuscript.

\end{document}